\documentclass[a4paper]{jpconf}
\usepackage{graphicx}

\def\lesssim{\mathrel{\hbox{\rlap{\hbox{\lower3.5pt\hbox{$\sim$}}}\hbox{$<$}}}}
\def\gtrsim{\mathrel{\hbox{\rlap{\hbox{\lower3.5pt\hbox{$\sim$}}}\hbox{$>$}}}}

\begin{document}
\title{Origin of ultra-high energy cosmic rays in the era of Auger and Telescope Array}

\author{Susumu Inoue}

\address{National Astronomical Observatory of Japan, 2-21-1 Osawa, Mitaka, Tokyo 181-8588, Japan
\footnote{Present address: Department of Physics, Kyoto University,
Oiwake-cho, Kitashirakawa, Sakyo-ku, Kyoto 606-8502, Japan; E-mail: inoue@tap.scphys.kyoto-u.ac.jp}}

\ead{inoue@th.nao.ac.jp}

\begin{abstract}
The origin of ultra-high energy cosmic rays is discussed
in light of the latest observational results from the Pierre Auger Observatory,
highlighting potential astrophysical sources such as
active galactic nuclei, gamma-ray bursts, and clusters of galaxies.
Key issues include their energy budget,
the acceleration and escape of protons and nuclei,
and their propagation in extragalactic radiation and magnetic fields.
We briefly address the prospects for Telescope Array and future facilities such as JEM-EUSO,
and also emphasize the importance of multi-messenger X-ray and gamma-ray signatures
in addition to neutrinos as diagnostic tools for source identification.
\end{abstract}

\section{Introduction}
\label{sec:intro}
Several decades after their discovery,
the origin of ultra-high energy cosmic rays (UHECRs),
cosmic particles with energies $10^{18}$-$10^{20}$ eV and above,
continue to be one of the biggest mysteries in astrophysics \cite{nw00}.
During the last few years,
the field has witnessed revolutionary advances,
thanks to the advent of new generation, hybrid detector facilities
spearheaded by the Pierre Auger Observatory in the southern hemisphere \cite{wat07}.
The Telescope Array \cite{fuk07} has begun operation in the northern hemisphere
and we eagerly await their data.
Expectations are already high for future projects
such as Auger North \cite{nit07}, or the Extreme Universe Space Observatory on the Japanese Experiment Module
(JEM-EUSO), to be deployed on the International Space Station \cite{tes07}.
This article, by no means a thorough review,
discusses selected topics in the rapidly developing field of UHECRs,
focusing on the theory of potential astrophysical sources
confronted with the latest observational results.
Some of the issues below are covered in more detail in earlier articles \cite{ino07}.

\section{General issues on propagation and recent observational results}
\label{sec:prop}

We first touch upon some general issues concerning the propagation of UHECRs 
(see e.g. \cite{bs00,pc04} for more details).
If UHECRs are protons of extragalactic origin,
photopion interactions with cosmic microwave background (CMB) photons
must induce severe energy losses at $\gtrsim 6 \times 10^{19}$ eV
for propagation lengths $\gtrsim 100$ Mpc.
Unless the sources lie much nearer,
a spectral steepening (Greisen-Zatsepin-Kuzmin or GZK cutoff \cite{gzk66})
is expected above these energies.
The Pierre Auger Observatory \cite{yam07}
as well as the HiRes experiment \cite{abb07}
have recently demonstrated the existence of a strong spectral steepening
at just this energy with high statistical significance,
although at present it cannot be excluded that
this represents instead some limit to acceleration at the sources.

The absence of super-GZK events reduces the motivation for the so-called top-down scenarios,
in which UHECRs arise as decay products of higher energy entities
envisaged in non-standard particle physics models,
such as topological defects, superheavy dark matter, etc. \cite{bs00}.
Moreover, they are seriously constrained by the tight limits on the photon fraction
from the latest Auger measurements \cite{sem07}.
In the following, we concentrate on the bottom-up, astrophysical scenarios.

A spectral steepening analogous to that for protons can also result if the highest energy UHECRs
are composed mainly of heavy nuclei such as iron.
For nuclei, the dominant energy loss process above $10^{19}$ eV during intergalactic propagation
is photodisintegration and pair production interactions with photons
of the far-infrared background (FIRB) and the CMB \cite{psb76}.
Based on recent determinations of the FIRB,
the energy loss length for iron nuclei at $10^{20}$ eV is $\gtrsim 300$ Mpc, 
somewhat larger than that for protons \cite{ss99}.
Observationally, the nuclear composition of UHECRs,
particularly at the highest energies, is quite uncertain.
Fluorescence measurements of shower elongation rates by the HiRes experiment
have indicated that the composition changes from heavy-dominated below $10^{18}$ eV
(presumably of Galactic origin) to light-dominated above this energy \cite{abb05}.
However, the latest results from Auger, based on hybrid events with higher precision
and considerably higher statistics, instead reveal an intermediate composition at all energies
up to $4 \times 10^{19}$ eV,
with hints of a trend toward a heavier composition at the highest energies \cite{ung07}.
A caveat is that
systematic uncertainties in interaction models and atmospheric optical attenuation
are still quite large \cite{wat04}.
These issues are also critical for the interpretation of the ankle feature in the spectrum at $\sim 3 \times 10^{18}$ eV
\cite{bgg06,all05}.
The Telescope Array should provide key complementary information in this regard,
since its scintillator surface detectors are relatively more sensitive
to the electromagnetic component of the air shower than the muon component,
in comparison with the water Cherenkov tanks of Auger.

Another unavoidable aspect
is deflection by Galactic and extragalactic magnetic fields (EGMF),
which affect the arrival directions of UHECRs
and can also act to lengthen their effective propagation distance.
The strength and distribution of magnetic fields in the intergalactic medium
as well as at high latitudes in the Galaxy are very poorly known, both observationally and theoretically.
Faraday rotation measurements of distant radio sources
give only upper limits in the nanogauss range for intergalactic fields on average,
subject to assumptions on the field coherence scale and ionized gas distribution \cite{kro94}.
Realistically, whatever their origin, EGMF can be expected
to have some correlation with the distribution of large scale structure,
and various EGMF models that account for this have been
introduced to assess their effects on UHECR propagation \cite{sme03}.
The effect of Galactic magnetic fields (GMF) is also model-dependent \cite{kst07,ts07}.

The Auger collaboration has recently announced statistically significant
evidence for UHECR anisotropy \cite{aug07}.
They find a $\sim 3\sigma$ correlation
between the arrival directions of 20 UHECR events above $5.6 \times 10^{19}$ eV,
and the positions of active galactic nuclei (AGNs) in the Veron-Cetty and Veron catalog
with distances $<$75 Mpc, to within $\sim 3$ degrees.
Some robust implications of this result are that:
1) the UHECR sky is not isotropic,
2) the sources of UHECRs are extragalactic and trace large-scale structure on scales of order 100 Mpc,
ruling out Galactic sources,
and 3) deflections of UHECRs in the GMF and EMGF are not so severe as to isotropize the arrival directions,
so that charged particle astronomy is possible.

Not much more can be said with confidence, however.
A correlation with the global distribution of AGNs does not necessarily mean that they are the true sources;
any other type of source that follows the distribution of large-scale structure,
such as gamma-ray bursts (GRBs) or clusters of galaxies, may also be feasible.
Inferences on the source density or UHECR composition from the current anisotropy results
are dependent on assumptions regarding the EGMF and GMF (see below).
In addition, the Veron-Cetty catalog is known to be highly non-uniform and incomplete,
and further studies with higher statistics utilizing more suitable catalogs are warranted
(see e.g. \cite{kas08} for recent studies in this direction).
Since the northern hemisphere should be less affected by the GMF,
data from the Telescope Array will be crucial, and all-sky observations by JEM-EUSO even better.
We note that the AGN correlation results have not been confirmed by the northern HiRes,
although their statistics is smaller \cite{abb08}.
At present, the true identity of the UHECR sources is still wide open.

\section{Acceleration and energetics in astrophysical sources of UHECRs}
\label{sec:source}

A minimum requirement for astrophysical sources of UHECRs
is the ability to magnetically confine particles of the requisite energies.
For particles with energy $E$ and charge $Z$, this implies the Hillas condition
$(R/{\rm pc}) (B/{\rm 1 G}) \gtrsim (E/{\rm 10^{20} eV})/Z$
between the system's size $R$ and magnetic field $B$ \cite{hil84}.
Note that this can also be rewritten as a lower limit on the source power,
assuming that the magnetic field carries a fixed fraction of the outflowing energy density \cite{nma95,bla99}.
Only a few types of objects are known to meet this criterion,
among them the jets of radio-loud AGNs,
GRBs, and clusters of galaxies.
Below we focus on these three
as representative types of potential UHECR sources
(see \cite{hil84} for other possibilities).
The actual maximum energy attainable
under different circumstances must be evaluated
by comparing the timescales for particle acceleration, often
that for the first-order Fermi mechanism in shocks,
against the timescales for limiting processes such as
source lifetime, particle escape, adiabatic or radiative energy loss, etc.

\begin{figure}[h]
\includegraphics[width=20pc]{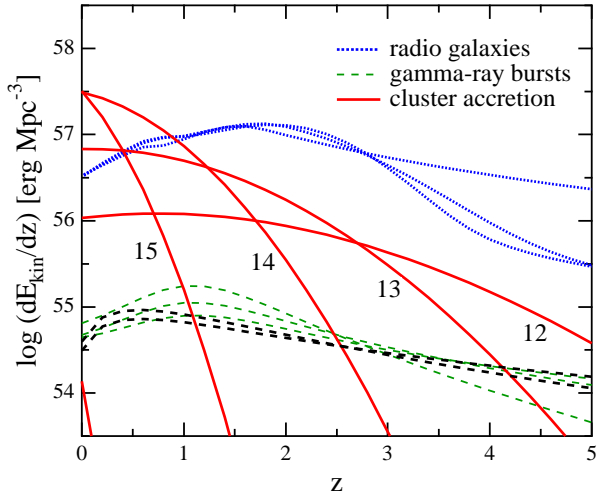}\hspace{2pc}
\begin{minipage}[b]{14pc}
\caption{
Energy budget of candidate UHECR sources:
AGN jets (dotted), GRBs (dashed; thin for BATSE, thick for Swift) and cluster accretion shocks (solid),
the latter separately for each $\log M$ as labelled. See text for more details.
\label{fig:ene}}
\end{minipage}
\end{figure}

Equally important is the available energy budget.
Fig.\ref{fig:ene} shows estimates of the kinetic energy output averaged over the universe
as a function of redshift $z$ due to AGN jets, GRB explosions and accretion onto clusters,
which should be roughly proportional to their cosmic ray output.
The plotted quantity is differential per unit $z$,
$dE_{\rm kin}/dz = (dt/dz) \int L (dn/dL) dL$,
where $L$ is the kinetic luminosity per object
and $dn/dL$ is the $z$-dependent luminosity function,
with cosmological parameters $h$=0.7, $\Omega_m$=0.3 and $\Omega_\Lambda$=0.7.
For AGN jets, we have made use of the observed radio luminosity function
along with the observed correlation between the radio and jet kinetic luminosities
of radio galaxies \cite{is01}.
GRBs were assumed to occur each with kinetic energy $E_{\rm GRB}=10^{54}$ erg
at a rate that follows the star formation history
and matches the $\log N$-$\log S$ distribution observed by BATSE \cite{pm01},
the estimate being roughly independent of the beaming factor.
Evaluations based on the more recent, post-Swift $z$-distribution \cite{ld07} are also shown.
The different curves for AGNs and GRBs correspond to
different evolutionary assumptions at the highest $z$,
with only small differences at low $z$.
For clusters with mass $M$,
the rate of gas kinetic energy dissipation through accretion shocks can be estimated as
$L_{\rm acc} \simeq 9 \times 10^{45} (M/{10^{15} M_\odot})^{5/3}$ erg/s \cite{kwl04,ias05}.
This can be combined with the Press-Schechter mass function
to evaluate $dE_{\rm kin}/dz$ for clusters of different $M$.
Note that due to the hierarchical nature of structure formation
together with the nonlinear nature of gravity,
the maximum is reached at $z=0$.

The results at low $z$ can be compared with the observed energy density of UHECRs,
$\simeq 10^{-19} {\rm erg \ cm^{-3}} \simeq 3 \times 10^{54} {\rm erg \ Mpc^{-3}}$ above $10^{19}$ eV.
Considering further factors for energy loss during propagation,
it is apparent that whereas AGN jets and cluster accretion shocks
have reasonable margins to accommodate the energetics of UHECRs,
GRBs, with a substantially smaller energy budget,
require a high efficiency of energy conversion into UHECRs.

\section{Active galactic nuclei}
\label{sec:agn}

AGNs are luminous active objects in the centers of some galaxies
that are powered by accretion of matter onto supermassive black holes \cite{kro98}.
Roughly 10 \% of them are known to be radio-loud,
ejecting powerful, relativistic jets from the nucleus out into its surroundings,
but the majority of AGNs are of the radio-quiet type with no strong jets.
In contrast to the former, the latter are considered unlikely to be UHECR sources
due to the severe radiative losses expected near the nucleus \cite{sp94},
and the absence of nonthermal emission components.

Radio-loud AGNs themselves come in two basic types \cite{up05}.
In FR II objects with radio (or jet kinetic) power above a critical value,
the jets remain relativistic out to large scales and end in luminous hot spots,
termination shocks where the jets are strongly decelerated upon impacting the intergalactic medium.
If such jets are oriented close to the line of sight,
they would appear as GeV-bright, flat spectrum radio quasars.
Contrastingly, the jets of FR I objects with subcritical power
are initially relativistic but is believed to decelerate gradually
as they propagate out of their host galaxies.
Their on-axis counterparts are the TeV-bright, BL Lac objects.

Different locations along AGN jets
can be UHECR production sites,
since one expects $B \propto R^{-1}$ under the naive assumption
that the ratio of magnetic to kinetic energy is constant.
For both FR II and FR I objects, 
a candidate is the inner jet region
with $R \sim 10^{16}$-$10^{17}$ cm and $B \sim 0.1$-$1$ G,
known through observations of blazars
to be a site of intense particle acceleration, perhaps due to internal shocks.
Estimates assuming Bohm-type acceleration
(i.e. acceleration time comparable to gyration time in fully turbulent magnetic fields)
show that the maximum proton energy should be limited
by photopion interactions with low frequency internal radiation
to somewhat below $10^{20}$ eV \cite{mpr01}.
However, plausible models of broadband blazar emission based on electron acceleration
generally indicate that the plasma conditions in inner jets are far from the Bohm limit \cite{it96}.
Furthermore, conversion to neutrons may be necessary
for the particles to escape the jet without suffering adiabatic expansion losses
and contribute to UHECRs by decaying back to protons outside.
This may entail rather finely-tuned conditions,
as photopion interactions must be efficient to yield enough neutrons,
yet cannot be excessively so to avoid adverse energy losses.
A less problematic site may be the hot spots of FR II radio galaxies
with $R \sim 10^{21}$ cm and $B \sim 1$ mG,
where the maximum energy may exceed $10^{20}$ eV, limited by escape \cite{bs87}.
However, note that these particles must further traverse
the extensive cocoon of shocked, magnetized jet material
in order to completely escape the system,
an issue that has not been examined in detail.
A further possibility is acceleration by the bow shocks
being driven into the external gas by the expansion of the cocoon,
if the ambient magnetic field can be sufficiently amplified \cite{nma95}.
In view of the Auger composition results \cite{ung07},
another interesting question is whether heavy nuclei such as iron
can be accelerated and survive photodisintegration in these environments \cite{der07}.

In order to verify an AGN origin,
detailed analysis of observed UHECR arrival directions
and cross correlations with known source positions is undoubtedly essential,
and the recent Auger results are an important step in this direction.
However, it is not a trivial task,
given the uncertainties in the intervening GMF and EGMF and the UHECR composition (Sec. \ref{sec:prop}).
In addition to deviations in the arrival directions, 
deflections in the EGMF during propagation also entail significant delays
in the arrival times of UHECRs, which can be of order $10^7$ yr or more for distances 100 Mpc,
depending on the properties of the EGMF \cite{wm96}.
This is in fact comparable to or longer than the typical lifetime or duration of AGN activity,
particularly for radio-loud AGNs \cite{mar03},
so that they may effectively behave as transient sources, in a way similar to GRBs.
Before the UHECRs emitted by an AGN reaches the observer,
the object may significantly evolve in its radio power, or possibly shut off its activity altogether.
Another implication is that an effectively larger number of sources contribute to small scale clustering,
underscoring the need for high event statistics as might be provided by JEM-EUSO or Auger North+South.

Because of these concerns,
it is also highly desirable to have some means to pinpoint individual sources
through characteristic, UHECR-induced signatures of secondary neutral radiation.
Neutrinos are ideal in providing an unambiguous earmark of high energy hadrons \cite{hh02}.
However, even with km$^{3}$ detector facilities such as IceCube or KM3NeT,
it may not be easy to study individual AGNs in sufficient detail.
Thus, distinctive electromagnetic signals will also be extremely valuable.
For radio galaxy hot spots,
synchrotron emission from UHE protons can produce nonthermal X-rays
that could be distinguished from other processes such as electron inverse Compton
through multiwavelength observations \cite{aha02}.
If the medium surrounding the source is sufficiently magnetized,
UHE protons propagating
in the source's vicinity can lead to diffuse gamma-ray emission from photomeson-triggered cascades
that may be detectable by current and upcoming instruments \cite{ga05}.

\section{Gamma-ray bursts}
\label{sec:grb}

GRBs are explosive phenomena that are believed to arise in a fraction of stellar collapse events,
characterized by highly variable, ultrarelativistic outflows lasting for $\sim 10-100$ s \cite{mes06}.
As with AGN jets, these outflows can be host to different sites for UHECR acceleration.
Potential locales include internal shocks, external reverse shocks, and external forward shocks,
believed to be the emission sites of the prompt X-rays and gamma-rays,
optical flash and radio flare, and the radio to X-ray afterglow, respectively \cite{wax95}.
The external forward shock may possibly be disfavored due to the ultrarelativistic velocity
and weak upstream magnetic field \cite{ga99}, but this is controversial \cite{der02}.
For the mildly relativistic internal and external reverse shocks,
a different problem is that for the particles to escape the acceleration site without significant losses,
neutron conversion may be required, as with AGN inner jet regions.

A crucial distinguishing feature of UHECRs from GRBs
is the narrow dispersion in CR energy expected for individual sources,
due to the time delay during propagation in EGMF \cite{wm96}.
Clear detections of this effect demand substantial event statistics
that may only be achievable with future facilities.
Even then, it would only demonstrate that UHECRs come from bursting sources,
and their GRB origin will remain ambiguous.
Thus, multi-messenger photon and neutrino signatures are particularly essential
in the case of GRBs.

Identification of high energy neutrino signals from GRBs are facilitated through time coincidence,
but studying individual bursts in detail may be difficult.
Thus, photon signatures of UHECR production will also be crucial \cite{da06}.
Detailed studies of this issue were conducted for the GeV-TeV emission from internal shocks,
utilizing a comprehensive Monte Carlo code that includes a wide variety of physical processes
related to high energy protons and electrons \cite{ai07,aim08}.
Besides inverse Compton emission from primary electrons,
interesting proton-induced components can become clearly visible,
such as synchrotron emission from high energy protons and muons,
as well as emission from secondary pair cascades injected by photomeson interactions,
serving as unique signatures of ultra-high-energy protons.

\begin{figure}[h]
\begin{center}
\includegraphics[width=17pc]{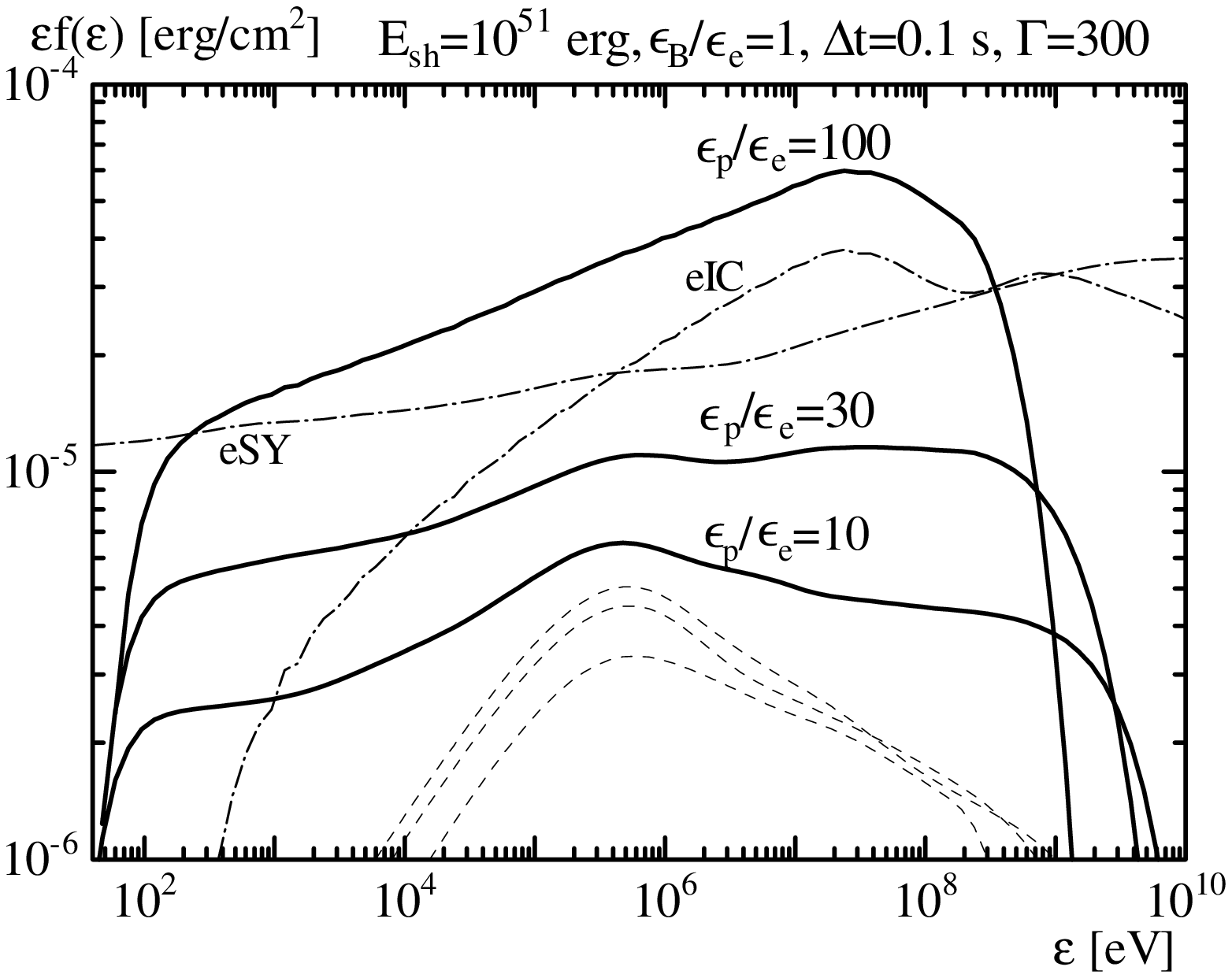}\hspace{2pc}
\includegraphics[width=17pc]{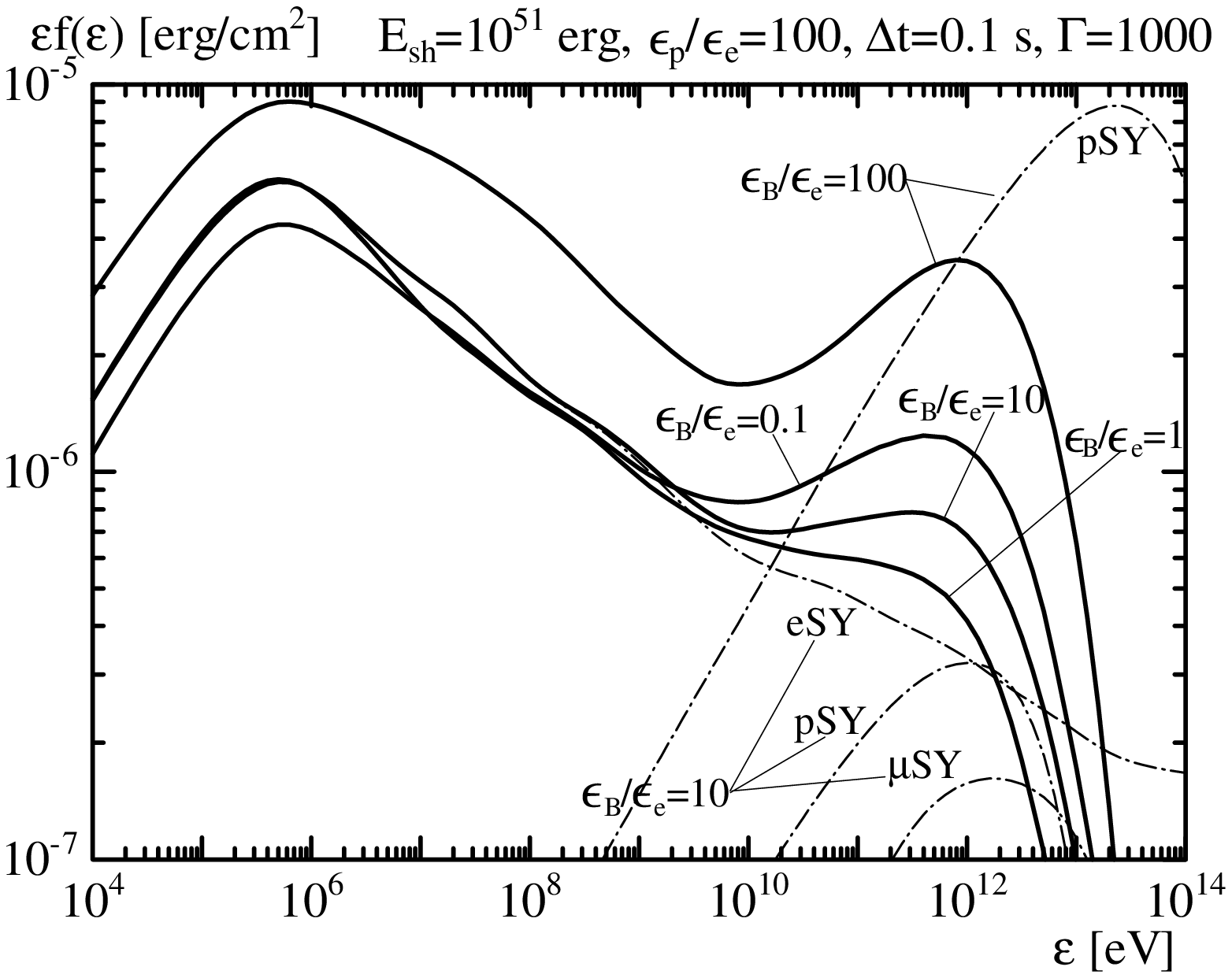}
\end{center}
\caption{
Prompt emission spectra from internal shocks in proton-dominated GRBs (see \cite{aim08} for more details).
(Left) 
For $\Gamma=300$ and different proton/electron ratios.
Solid curves are the total emission, and dot-dashed curves denote separately
electron synchrotron (eSY) and electron inverse Compton (eIC) components
without $\gamma \gamma$ absorption effects.
Dashed curves are primary electron components only.
(Right) Similar to left panel, except for $\Gamma=1000$ and different magnetic fields.
Dot-dashed curves denote separately
proton synchrotron (pSY) and muon synchrotron ($\mu$SY) components.
\label{fig:grbspec}}
\end{figure}

Particularly interesting are proton-dominated GRBs
that contain a significantly larger amount of energy
in accelerated protons compared to accelerated electrons \cite{aim08}.
Such conditions are motivated not only by the physics of particle acceleration in collisionless shocks,
but also by the latest measurements of the GRB $z$-distribution.
Post-Swift estimates of the local GRB rate \cite{ld07,gp07}
indicate that the proton energy content per burst may need to be substantially higher than previously thought
in order for them to remain viable as UHECR sources.
Unique UHECR-induced GeV-TeV components are then expected,
such as GeV peaks, UV-X-ray excesses and luminous TeV bumps (Fig.\ref{fig:grbspec}).

Note that the conditions favorable for GeV-TeV emission may also imply
efficient UHECR acceleration and escape, but not necessarily strong neutrino emission \cite{ai07}.
The observational prospects are promising for GLAST,
atmospheric Cherenkov telescopes such as MAGIC (II), H.E.S.S. (II), VERITAS and CANGAROO III,
as well as wide-field surface detector facilities,
which should test these expectations and provide important information on the physical conditions in GRB outflows.

A different class of low luminosity GRBs have been proposed to be more important
than high luminosity GRBs for the integrated energy budget,
and hence potentially more relevant as sources of UHECRs \cite{mur06}.
As with AGN jets, the acceleration and survival of heavy nuclei in
in the environments of both high and luminosity GRBs is an important question \cite{anc07}.

\section{Cluster accretion shocks}
\label{sec:clus}

Clusters of galaxies are the largest gravitationally bound systems in the universe,
composed of stars, hot gas and dark matter.
According to the currently favored picture of hierarchical structure formation in the CDM cosmology,
they are the latest objects to form, and 
should be surrounded by strong accretion shocks
as a consequence of continuing infall of dark matter and baryonic gas \cite{min00}.
Such shocks should be interesting sites of particle acceleration
and have been proposed as sources of UHECRs \cite{nma95}.
However, estimates of the maximum energy $E_{\max}$ for protons
seem to fall short of $10^{20}$ eV by 1-2 orders of magnitude \cite{krj96,ias05}.

\begin{figure}[h]
\includegraphics[width=18pc]{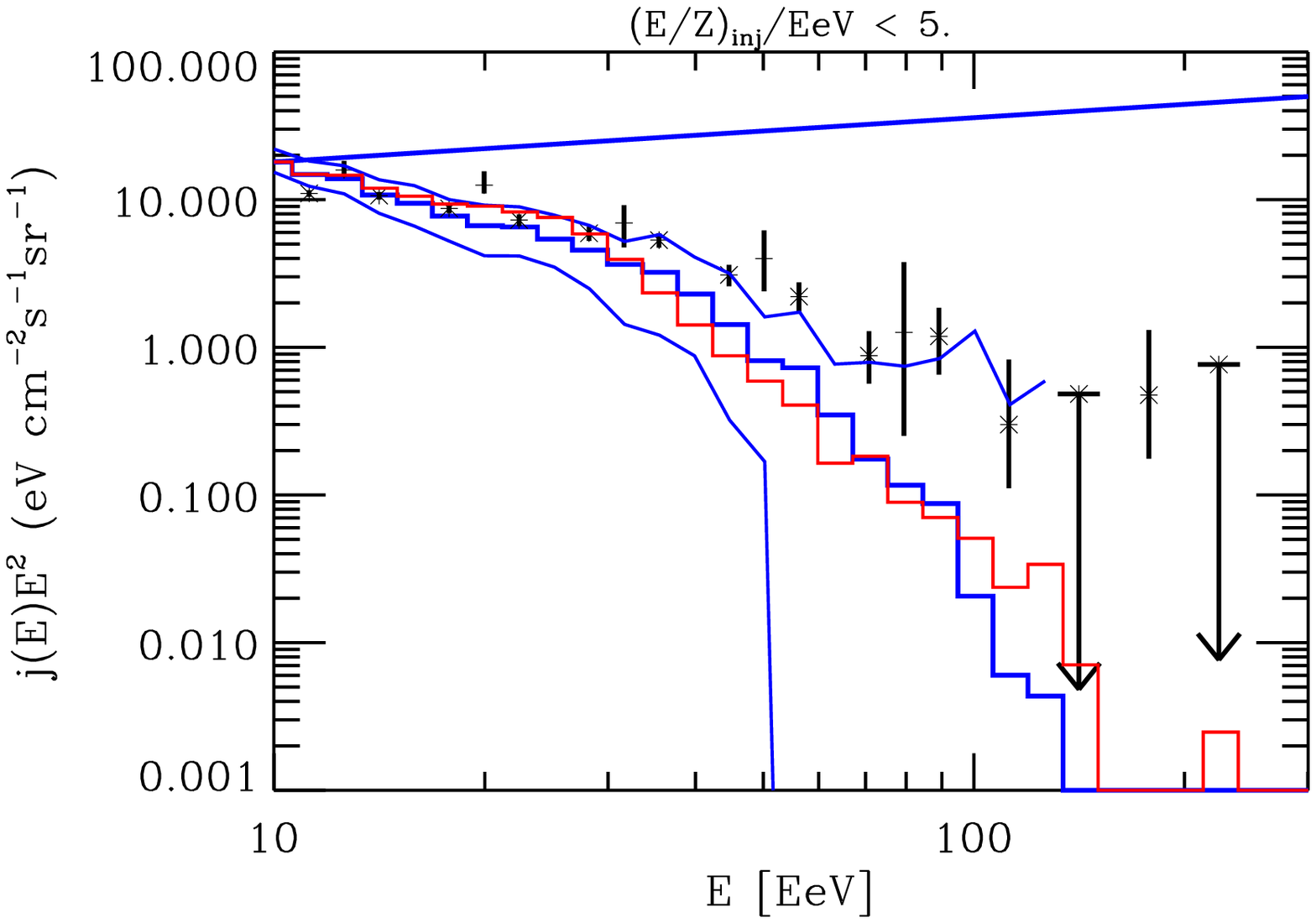}
\includegraphics[width=18pc]{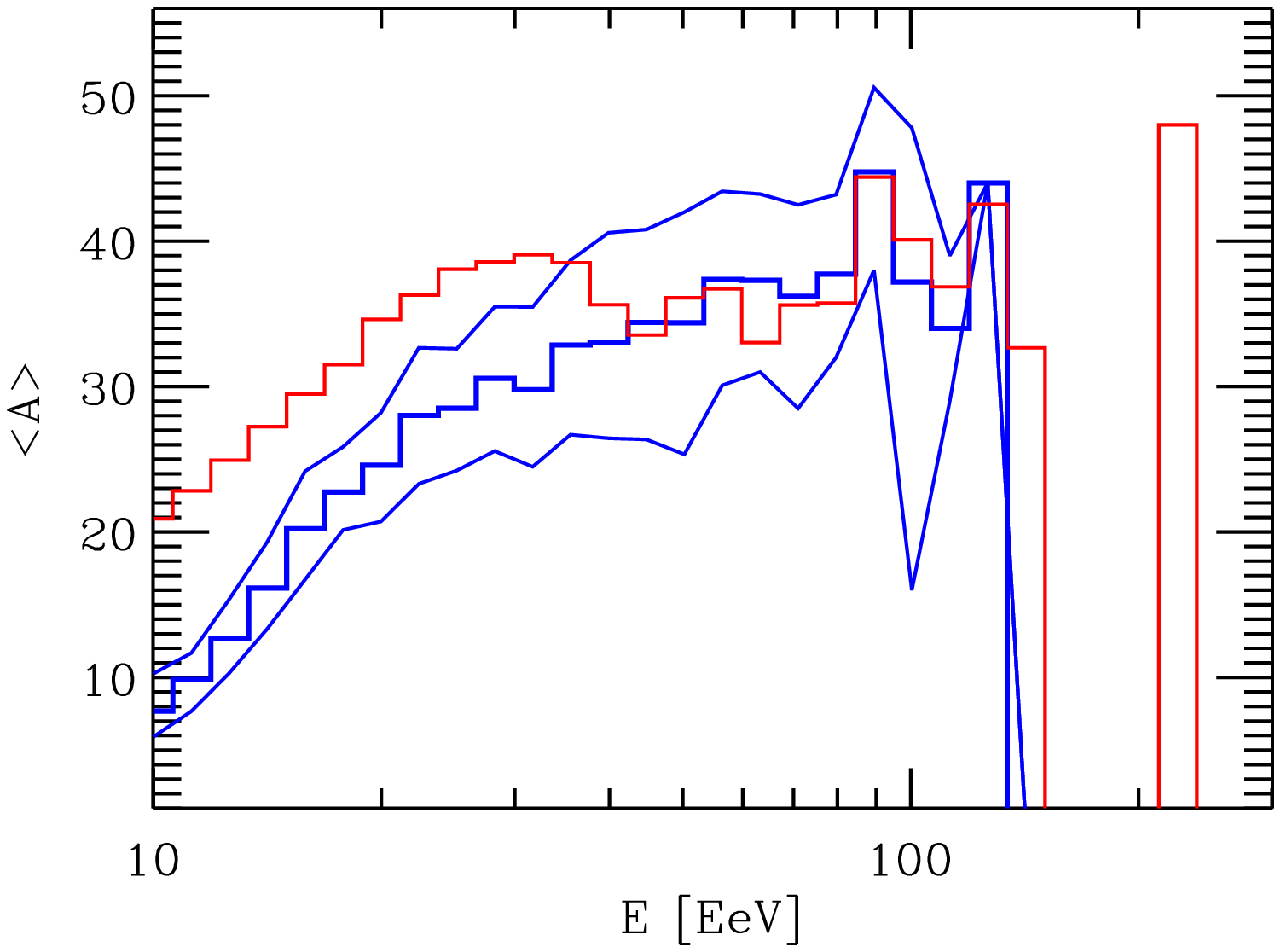}
\caption{Observed UHECR spectrum (left) and mean mass composition (right)
versus energy $E$ (1 EeV $\equiv 10^{18}$ eV)
from cluster accretion shocks for $\alpha=1.7$ and $\beta=0.5$ \cite{isma07}.
Compared are current data for
HiRes \cite{abb04} (bars) and Auger \cite{yam07} (asterisks).
The histograms are the average over different model realizations
for the cases with (thick) and without (thin) EGMF,
and the thin curves outline the median deviations due to cosmic variance for the former case only.
The straight line in the top panel denotes the injection spectrum.
\label{fig:speccomp}}
\end{figure}

We have shown that invoking UHECR nuclei in cluster accretion shocks
may offer a possible solution \cite{isma07}.
Heavy nuclei with higher $Z$ have correspondingly shorter acceleration time
so that Fe may be accelerated up to $10^{20}$ eV in the same shock conditions,
notwithstanding energy losses by photodisintegration with the FIRB and CMB.
With reasonable assumptions regarding
the number density and CR power of the sources,
detailed propagation calculations demonstrate
that the model compares favorably with the existing observations,
as long as the source composition of Fe is similar to that of Galactic CRs (Fig.\ref{fig:speccomp}).
See also \cite{ari07} for related studies on UHECR nuclei propagation.

The mass composition at $\lesssim 3 \times 10^{19}$ eV should be predominantly light
and consistent with HiRes reports \cite{abb05}.
The rapid increase of the average mass at higher energies
is a clear prediction of the scenario,
in line with the latest Auger elongation measurements \cite{ung07},
and to be tested further by new generation facilities including Telescope Array and JEM-EUSO.

Despite the relative rarity of massive clusters in the local universe,
strong deflections of the highly charged nuclei in EGMF
allow consistency with the global isotropy.
On the other hand, with a sufficient number of accumulated events,
anisotropies toward a small number of individual sources should appear,
although this expectation is subject to uncertainties in the EGMF and GMF.
Whether it is consistent with the latest Auger anisotropy results calls for further investigation.
It could possibly be the case, including the observed deficit of events from the Virgo direction,
if the GMF is of the plausible, bisymmetric spiral type,
in which deflection angles can be minimal
in selected regions of the sky around the supergalactic plane,
particularly around Centaurus \cite{ts07}
(see also \cite{wib07} for related discussions).
Note that the Centaurus cluster at distance $\sim$50 Mpc
is one of most prominent clusters in the local universe,
somewhat more massive than Virgo \cite{rb02}.

We mention that the escape of CRs from clusters
can be mediated by diffusion in directions away from the filaments,
or perhaps by advection in partial outflows during merging events.
Energy-dependent escape upstream of the shock may also be possible.
An aspect of this scenario that warrants further study is the spectral domain $< 10^{19}$ eV
and the implications for the Galactic-extragalactic transition region.
Alternatively, provided that EGMF horizon effects are effective $\lesssim 10^{17}$ eV \cite{kl08},
cluster accretion shocks may give an appreciable contribution to CRs between the second knee and the ankle,
in which case they may also be interesting as high energy neutrino sources \cite{mi08}.

Finally, we may look forward to very unique signatures in X-rays and gamma-rays.
Protons accelerated to $10^{18}$-$10^{19}$ eV in cluster accretion shocks
should efficiently channel energy into pairs of energy $10^{15}$-$10^{16}$ eV
through interactions with the CMB,
which then emit synchrotron radiation peaking in hard X-rays
and inverse Compton radiation in TeV gamma-rays \cite{ias05}.
The detection prospects are promising
for Cherenkov telescopes such as HESS,
and hard X-ray observatories such as Suzaku and the future NeXT mission.
Photopair production by nuclei may also be efficient
and induce further interesting signals that are worth investigating.

\section{Conclusion and outlook}
\label{sec:conc}
Besides AGNs, GRBs and clusters, other proposals for the astrophysical origin of UHECRs
not covered in this brief review
include starburst galaxies \cite{anc99},
extragalactic magnetars \cite{aro03},
dormant black holes \cite{bol99}, etc. (see \cite{hil84}).
The true source of UHECRs could be one or more of these;
alternatively, we should not discount the possibility that it is actually none of them,
and the truth something totally unexpected.

In this regard, we may recall the history of GRB research.
Before the launch of BATSE in 1991, the sample of GRBs numbered $\lesssim 200$,
with no clear anisotropy observed in the sky distribution.
However, most believed, based on theoretical plausibility,
that they were generated by Galactic neutron stars,
while a very small minority advocated neutron star mergers at cosmological distances.
After several years and a sample of more than 3000 bursts,
the currently most favored scenario for the progenitors of long GRBs
is related to the collapse of massive stars \cite{woo93}, a concept that didn't exist before BATSE.
The lesson is that we should be cautious about jumping to conclusions
on the origin of UHECRs based on just 20 observed events
and our limited theoretical understanding of cosmic objects.

Regardless of theoretical prejudice,
the quest for the solution of the UHECR mystery is progressing in earnest
with ongoing measurements by Auger and Telescope Array.
An order of magnitude greater statistics for the highest energy events expected
from JEM-EUSO or Auger North+South,
and the combined effort of upcoming neutrino, X-ray and gamma-ray observations
is expected to lead us toward the true answer in the near future.

\ack
The author is thankful for past and ongoing collaborations with
F. Aharonian, E. Armengaud, K. Asano, N. Kawakatu, P. Meszaros, F. Miniati, K. Murase, K. Sato,
G. Sigl, N. Sugiyama, H. Takami and T. Yamamoto.

\end{document}